\documentclass[twocolumn,showpacs,prl]{revtex4}

\usepackage{graphicx}
\begin{document}
%
\title{Exact Results for the Reactivity of a Single-File System}
\author{A. P. J. Jansen}
 \altaffiliation{Department of Chemical Engineering, ST/SKA}
 \email{tgtatj@chem.tue.nl}
\author{S. V. Nedea}
 \altaffiliation{Department of Mathematics and Computing Science}
\author{J. J. Lukkien}
 \altaffiliation{Department of Mathematics and Computing Science}
\affiliation{Schuit Institute of Catalysis,
Eindhoven University of Technology,\\
P. O. Box 513,
5600 MB Eindhoven,
The Netherlands}
\date{\today}
\begin{abstract}
  We derive analytical expressions for the reactivity of a Single-File
  System with fast diffusion and adsorption and desorption at one end.
  If the conversion reaction is fast, then the reactivity depends only
  very weakly on the system size, and the conversion is about 100\%. If
  the reaction is slow, then the reactivity becomes proportional to the
  system size, the loading, and the reaction rate constant. If the
  system size increases the reactivity goes to the geometric mean of the
  reaction rate constant and the rate of adsorption and desorption. For
  large systems the number of nonconverted particles decreases
  exponentially with distance from the adsorption/desorption end.
\end{abstract}
\maketitle
%
%

Zeolitic structures like mordenite, ZSM-22, and ${\rm AlPO}_4$-o5 are
industrially immensely important. They are examples of single-file
systems (SFS).\cite{mei96} They have one-dimensional pores through which
molecules can diffuse, but with cross sections that are too small to
allow passing. As for many other 1D systems, this leads to very
interesting kinetic effects.\cite{pri97} Research on SFS zeolites has
focussed on the diffusion, because the mean-square displacement in a SFS
is proportional to the square root of time, and not proportional to
time.\cite{fed78,bei83,hah96} Molecular
Dynamics,\cite{kef95,hah96,sch00d} Dynamic Monte Carlo
(DMC),\cite{rod95,nel99,ned02} and reaction-diffusion
equations\cite{kri00} have mainly been used. Few studies have included
reactions.\cite{tsi91,kar92,ned02} We have argued that this is
unfortunate, because if diffusion is fast the temporal dependence of the
mean-square displacement is not very relevant, but there are still
effects from the non-passing of the particles in the pores.

In a previous paper we have investigated steady-state properties of a
SFS for different assumptions of the reactive site distribution using
DMC as well as analytical techniques.\cite{ned02} In the current paper
we derive exact results for the case that the entire pore is reactive.
These results give detailed insight in the relationships of the various
system parameters and the reactivity.

%
%

%
%
%

Our model consists of $S$ sites forming a one-dimensional finite chain.
Each site is numbered consecutively from 1 for the site on one end, to
$S$ for the site on the other end.  Each site is vacant or is occupied
by a particle. We have two types of particles: A and B.  An A can be
converted into a B on any site. Adsorption and desorption of particles
can only occur on site 1. Only A's are adsorbed, but both A's and B's
desorb. Both types of particle diffuse by making random hops to
neighboring sites if vacant.

The evolution of the system is described by a master
equation\cite{gel99,kam81}
\begin{equation}
  {dP_\alpha\over dt}
  =\sum_\beta\left[W_{\alpha\beta}P_\beta-W_{\beta\alpha}P_\alpha\right],
\end{equation}
where $\alpha$ and $\beta$ refer to the configuration of the adlayer (a
particular distribution of particle over the sites), the $P$'s are the
probabilities of the configurations, $t$ is real time, and the $W$'s are
constants that give the rates with which reactions change the
occupations of the sites. $W_{\alpha\beta}$ corresponds to the reaction
that changes $\beta$ into $\alpha$. The rate constants in our model are
$W_{\rm ads}$ for adsorption of an A onto site 1 if vacant, $W_{\rm
  des}$ for desorption of a particle from site 1, $W_{\rm rx}$ for the
conversion of an A into a B on any site, and $W_{\rm diff}$ for a hop of
a particle to a vacant neighboring site.
%
%
%

If the diffusion is infinitely fast, we can derive a simpler master
equation for the number of particles in the system.
\begin{eqnarray}
  {dP_N\over dt}&=&W_{\rm ads}\left[1-{N-1\over S}\right]P_{N-1}
                  -W_{\rm ads}\left[1-{N\over S}\right]P_N\nonumber\\
                &+&W_{\rm des}{N+1\over S}P_{N+1}
                  -W_{\rm des}{N\over S}P_N
\label{eq:MENpart}
\end{eqnarray}
where $P_N$ is the probability that there are $N$ particles in the
system. This is a master equation of a one-step Markov
process.\cite{kam81}
%
%
%

We are interested in the probability distribution $f_{MK}(t)$ that if
at time $t=0$ the number of particles is $K$, this number becomes $M$ for
the first time at time $t$ with $M<K$. If there are $N$ particles in the
system and at time $t=0$ a particle adsorbs, then $f_{N,N+1}$ is the
probability distribution for the time that this particle desorbs. This
is based on the property of a SFS that particles can not pass each
other. So $f_{N,N+1}$ is the probability distribution for the residence
time of a particle that adsorbs in a system with $N$ particles.

Let $P_{NK}$ be the solution of Eq.~(\ref{eq:MENpart}) with $N$ the
number of particles and the initial condition $P_{NK}(0)=\delta_{NK}$.
Let $Q_{NK}$ also be a solution with $Q_{NK}(0)=\delta_{NK}$, but now
for the master equation with an adsorbing boundary at $M$: i.e., we
remove the term in Eq.~(\ref{eq:MENpart}) that corresponds to an
adsorption then there are $M$ particles in the system. With $N\ge M$ we
have
\begin{equation}
  P_{NK}(t)=Q_{NK}(t)
            +\int_0^t\!\!dt^\prime P_{NM}(t-t^\prime)f_{MK}(t^\prime).
\end{equation}
This equation is called the renewal equation.\cite{kam81} If we take
$N=M$ in the renewal equation, then we have $Q_{MK}(t)=0$ by
definition. So we get the integral equation
\begin{equation}
  P_{MK}(t)=\int_0^t\!\!dt^\prime
  P_{MM}(t-t^\prime)f_{MK}(t^\prime)
\label{eq:reneq}
\end{equation}
for $f_{MK}$.
%
%

%
%
%

A particle that adsorbs, stays in the system for a period $t$, and then
desorbs, has a probability $\exp(-W_{\rm rx}t)$ that it is not converted
during the time it was in the system. Because the probability
distribution for the time that it stayed in the system is given by
$f_{N,N+1}(t)$, the probability that the particle desorbs without being
converted is
\begin{equation}
  \int_0^\infty\!\!dt\,\exp(-W_{\rm rx}t)f_{N,N+1}(t).
\label{eq:laplace}
\end{equation}
This is equal to $\hat f_{N,N+1}(W_{\rm rx})$ with $\hat f_{N,N+1}$ the
Laplace transform of $f_{N,N+1}$. The reason for using this Laplace
transform is that is that it is related to the Laplace transform of the
solution of the master equation through the renewal equation. Laplace
transforming Eq.~(\ref{eq:reneq}) yields
\begin{equation}
  \hat P_{MK}(s)=\hat P_{MM}(s)\hat f_{MK}(s).  
\end{equation}
If we write the master equation~(\ref{eq:MENpart}) in matrix-vector
notation as $\dot P={\bf W}P$, and Laplace transform it we get
\begin{equation}
  \sum_M(s\delta_{NM}-{\bf W}_{NM})\hat P_M(s)=P_N(0)
\end{equation}
With the initial condition for $P_{MK}$ this yields
\begin{equation}
  \hat P_{MK}(s)=[(s-{\bf W})^{-1}]_{MK},
\end{equation}
so that
\begin{equation}
  \hat f_{MK}(s)={[(s-{\bf W})^{-1}]_{MK}\over[(s-{\bf W})^{-1}]_{MM}}.
\label{eq:reci}
\end{equation}

We define the reactivity $B_{\rm prod}$ as the number of particles that
is being converted per unit time. To get $B_{\rm prod}$ we have to
multiply the probability that a particle is converted by the rate of
adsorption and make a weighted average over the number of particles in
the system. The probability that a particle is converted equals $1-\hat
f_{N,N+1}(W_{\rm rx})$, and the rate of adsorption equals $W_{\rm
  ads}(1-N/S)$. At steady state the number of particles in the system is
given by\cite{ned02}
\begin{equation}
  P_N^{(ss)}=\pmatrix{S\cr N\cr}
    \left[{W_{\rm des}\over W_{\rm ads}+W_{\rm des}}\right]^S
    \left[{W_{\rm ads}\over W_{\rm des}}\right]^N.
\label{eq:pocc}
\end{equation}
Combining this results in
\begin{equation}
  B_{\rm prod}=W_{\rm ads}\sum_{N=0}^{S-1}P_N^{(ss)}
    \left[1-{N\over S}\right]
    \left[1-\hat f_{N,N+1}(W_{\rm rx})\right].
\end{equation}

\begin{figure}
\includegraphics[width=0.9\hsize]{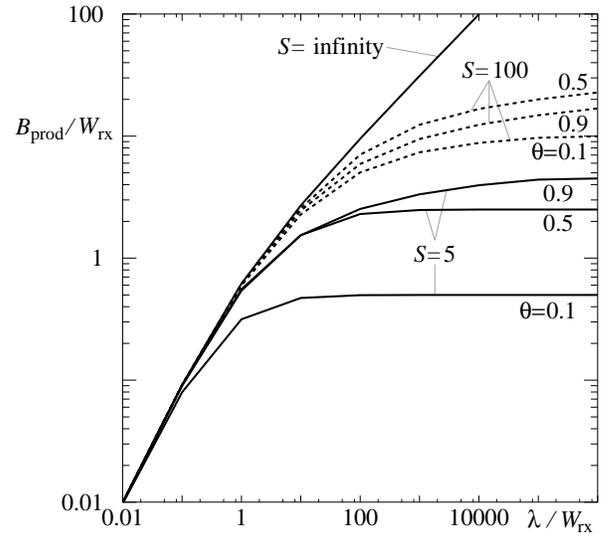}
\caption{Reactivity as a function of rate of adsorption and
  desorption. The reactivity and the rate of adsorption and desorption
  are scaled with the rate constant of the conversion.}
\label{fig:bprod}
\end{figure}

Figure~\ref{fig:bprod} shows this reactivity for two finite system
sizes and three loadings. The loading $\theta$ is defined as the
probability that a site is occupied at steady state. It's equal
to\cite{ned02}
\begin{equation}
  \theta={W_{\rm ads}\over W_{\rm ads}+W_{\rm des}}.
\end{equation}
The reactivity is compared to the (average) rate of adsorption and
desorption $\lambda$, which at steady state is given by
\begin{equation}
  \lambda=W_{\rm ads}(1-\theta)=W_{\rm des}\theta.
\end{equation}
If the rates of adsorption and desorption are small compared to the
reaction rate constant $W_{\rm rx}$, then $B_{\rm prod}\to\lambda$.
This means that almost every particle that enters the system is
converted. If the rates of adsorption and desorption are much larger,
then
\begin{equation}
  B_{\rm prod}\to S\theta W_{\rm rx}
\end{equation}
for $\lambda/W_{\rm rx}\to\infty$ and $S$ not too large.
Figure~\ref{fig:bprod} shows the $\theta$ dependence for small $S$ and
the $S$ dependence for small $\theta$. It does not show this $S$ and
$\theta$ dependence for $S$ and $\theta$ both large. $\lambda$ needs to
be much larger than the values in the figure for that. For $S=100$ we
see that the reactivity may be proportional to $\theta$ for $\theta\le
0.5$, but not around $\theta=0.9$. For the latter value the reactivity
is even smaller than for $\theta=0.5$.

In practice only $W_{\rm ads}$ can usually be varied independently,
either through the pressure or the concentration of A's outside the
system. The limit $W_{\rm ads}\to\infty$ corresponds to $\theta=1$ and
$\lambda=W_{\rm des}$. So we see that there generally is an upper bound
on $\lambda$.

Comparing $S=5$ and $S=100$ in Fig.~\ref{fig:bprod} for some finite
$\lambda$ shows that $B_{\rm prod}$ varies less with $\theta$ for larger
$S$. We can show that in the limit $S\to\infty$ the reactivity depends
only on $\lambda$ and $W_{\rm rx}$. If the system becomes infinitely
large, then the fluctuations become small with respect to the number of
sites. If the system is then at steady state we can write the master
equation as
\begin{equation}
  {dP_N\over dt}=\lambda(P_{N+1}+P_{N-1}-2P_N).
\end{equation}
Formally we can let the number of particles $N$ run from $-\infty$ to
$+\infty$. The matrix $s-{\bf W}$ can easily be diagonalized. The
eigenvalues are $s+2\lambda(1-\cos k)$ with $-\pi<k\le\pi$, and the
corresponding eigenvector has components $\exp[ikN]$. We can use this to
calculate elements of the inverse of the matrix $s-{\bf W}$.
\begin{eqnarray}
  [(s-{\bf W})^{-1}]_{NM}
  &=&{1\over 2\pi}\int_{-\pi}^\pi\!\!dk
     {\cos(\vert N-M\vert k)\over s+2\lambda(1-\cos k)}\nonumber\\
  &=&{1\over ys}\left[{y-1\over y+1}\right]^{\vert N-M\vert}.
\label{eq:invmat}
\end{eqnarray}
with
\begin{equation}
  y=\sqrt{1+4\lambda/s}.
\end{equation}
This then yields for the probability that a particle adsorbs and desorbs
at a later time without being converted the expression
\begin{equation}
  \hat f_{N,N+1}(W_{\rm rx})
  ={y-1\over y+1}.
\end{equation}
(Here and in the rest of the paper we set $s=W_{\rm rx}$ in the
definition of $y$.) For the reactivity $B_{\rm prod}$ we have
\begin{equation}
  B_{\rm prod}=\lambda[1-\hat f_{N,N+1}(W_{\rm rx})]
              ={1\over 2}W_{\rm rx}(y-1).
\end{equation}
We see that the reactivity no longer depends on the loading $\theta$ if
the system becomes large, but only on the reaction rate and on
$\lambda$. For small rates of adsorption and desorption we again find
$B_{\rm prod}\to\lambda$. If the rates of adsorption and desorption are
large, then
\begin{equation}
  B_{\rm prod}\to\left({\lambda W_{\rm rx}}\right)^{1/2}
\end{equation}
for $\lambda/W_{\rm rx}\to\infty$. The approach to the limit
$S\to\infty$ becomes very slow when $\lambda$ is large as can be seen in
Fig.~\ref{fig:bprod}.
%
%
%

The procedure for the first-passage problem above can also be used to
derive the A and B profiles: i.e., the distribution of the A's and B's
in the chain. We will first deal with the question what the probability
is that the $n$th particle, counting from site 1, is an A. Then we will
answer the question that site $m$ is occupied by an A. The answers to
the same questions about B's follow trivially from the ones for the A's.

We are only interested in the steady state. In this case we have
detailed balance: i.e., the number of transitions per unit time from $N$
to $N+1$ particles is equal to those of $N+1$ to $N$. This means that
for each sequence in the number of particles
\begin{equation}
   N_0
   \mathop{\longrightarrow}^0 N_1
   \mathop{\longrightarrow}^{\Delta t_1} N_2
   \mathop{\longrightarrow}^{\Delta t_2} \ldots
   \mathop{\longrightarrow}^{\Delta t_{T-1}} N_T=N_0,
\end{equation}
where after a time lapse $\Delta t_i$ the number of particles changes
from $N_i$ to $N_{i+1}$, there is another sequence
\begin{equation}
   N_0=N_T
   \mathop{\longrightarrow}^0 N_{T-1}
   \mathop{\longrightarrow}^{\Delta t_{T-1}} N_{T-2}
   \mathop{\longrightarrow}^{\Delta t_{T-2}} \ldots
   \mathop{\longrightarrow}^{\Delta t_1} N_0.
\end{equation}
The second sequence is the time reversed of the first one. Moreover,
both sequences are equally likely because of the detailed
balance. Consequently, the probability distribution that the number
of particles in the system is $K$ at time $t=0$ and $M$ with $M<K$ at
time $-t$ for the last time equals $f_{MK}(t)$. This means that if
there are $N$ particles in the system the probability distribution that
particle $n$ is in the system for a time $t$ equals $f_{N-n,N}(t)$. The
probability that that particle has not been converted is then $\hat
f_{N-n,N}(W_{\rm rx})$ following the reasoning after
Eq.~(\ref{eq:laplace}). This probability can be calculated using
Eq.~(\ref{eq:reci}).

For the profile we need the probability that site $m$ is occupied by
particle $n$. This probability is given by
\begin{equation}
  P_{\rm occ}(n,N;m,S)
  =\pmatrix{m-1\cr n-1\cr}
   \pmatrix{S-m\cr N-n\cr}/
   \pmatrix{S\cr N\cr}.
\end{equation}
The probability $\langle{\rm A}_m\rangle$ that site $m$ is occupied by a
particle that has not been converted is then given by
\begin{eqnarray}
  \langle{\rm A}_m\rangle
  &=&\sum_{N=0}^SP_N^{(ss)}\pmatrix{S\cr N\cr}^{-1}
\label{eq:Psite}\\
  &\times&\sum_{n=1}^m\pmatrix{m-1\cr n-1\cr}\pmatrix{S-m\cr N-n\cr}
  \hat f_{N-n,N}(W_{\rm rx})\nonumber
\end{eqnarray}
where the first summation averages over the number of particles in the
system and $P_N^{(ss)}$ is given by Eq.~(\ref{eq:pocc})

The expression above can be simplified and interpreted more readily for
an infinite system. With Eq.~(\ref{eq:invmat}) we have
\begin{equation}
  \hat f_{N-n,N}=\left({y-1\over y+1}\right)^n
\label{eq:Pconv}
\end{equation}
for the probability that particle $n$ has not been converted. We see
that this probability decreases exponentially.

The probability $P_{\rm occ}(n,N;m,S)$ becomes $P_{\rm occ}(n,m;\theta)$
with $\theta=N/S$ for $S\to\infty$. This limit of the combinatorial
factors yields
\begin{equation}
  P_{\rm occ}(n,m;\theta)=\pmatrix{m-1\cr n-1\cr}\theta^n(1-\theta)^{m-n}.
\end{equation}
Substituting this expression and (\ref{eq:Pconv}) in (\ref{eq:Psite})
yields
\begin{equation}
  \langle{\rm A}_m\rangle
  ={\theta(x-s)(x+s-2\theta s)^{m-1}\over(x+s)^m}.
\label{eq:prof}
\end{equation}
Again we find an exponential decrease but now with a characteristic
length of $\Delta=1/[\ln(y+1)-\ln(y+1-2\theta)]$. For high loadings
($\theta\to 1$) we find $\Delta\approx y/2$. For low loading ($\theta\to
0$) we get $\Delta\approx(y+1)/(2\theta)$. So the characteristic length
is larger for low loadings and higher rate of adsorption and
desorption.

\begin{figure}
\includegraphics[width=0.9\hsize]{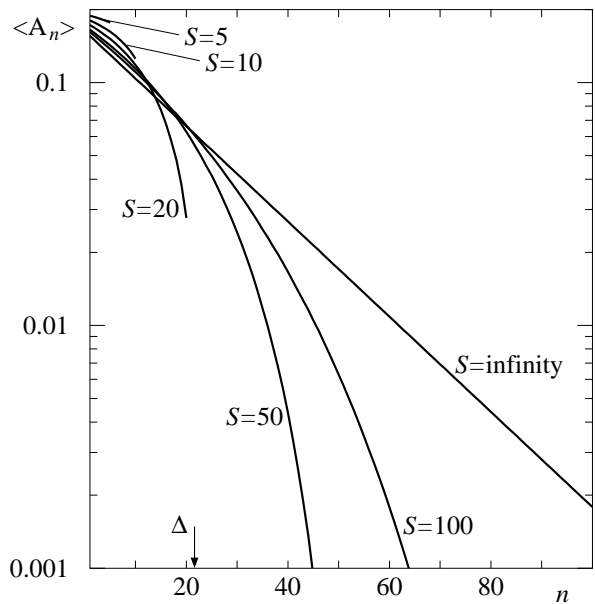}
\caption{The probabilities $\langle{\rm A}_n\rangle$ that a site is
  occupied by a non-converted particle as a function of site index $n$
  for rate of adsorption and desorption $\lambda=16$, loading
  $\theta=0.2$, and various system sizes $S$.}
\label{fig:prof}
\end{figure}

Figure~\ref{fig:prof} shows some typical profiles. The straight line
corresponds to the exponential decrease of Eq.~(\ref{eq:prof}). The
result for finite system sizes can be understood from the fact that
smaller systems are less reactive because there are fewer sites at which
conversion can take place. As the reactivity must be equal to the number
of desorbing converted particles, $W_{\rm des}\langle{\rm B}_1\rangle$
must be smaller for smaller $S$, and because $\langle{\rm
  B}_1\rangle=\theta-\langle{\rm A}_1\rangle$ the curves in
Fig.~\ref{fig:prof} must start out at larger values of $\langle{\rm
  A}_1\rangle$. The reactivity also equals $W_{\rm
  rx}\sum_{n=1}^S\langle{\rm A}_n\rangle$. This means that
$\sum_{n=1}^S\langle{\rm A}_n\rangle$ must be smaller for smaller $S$.
Because $\langle{\rm A}_1\rangle$ is larger for smaller $S$,
$\langle{\rm A}_n\rangle$ must decrease faster.

Figure~\ref{fig:prof} confirms the slow convergence of the system to the
limit $S\to\infty$. At $S=100$ the values of $\langle{\rm A}_n\rangle$
drop down to $2\cdot 10^{-5}$. At such low value one would not expect an
influence of these sites on the kinetics, but one clearly sees
differences between $S=100$ and $S\to\infty$ for all sites.

We would finally like to comment on three extensions of our model:
adsorption and desorption at both ends, finite diffusion, and other
reactions. A system with $2S$ sites and adsorption and desorption at
both ends has a reactivity that is twice the one of a system with $S$
sites and adsorption and desorption at just one end provided that
$S\gg\Delta$. DMC simulation show that if $S$ is near or smaller than
$\Delta$, then the system open at both ends is relatively more reactive.

Finite diffusion is expected to be equivalent to infinite diffusion as
long as it is much faster than the other reactions. If this is not the
case, then it may become a rate determining step. The system then becomes
transport limited. As a consequence the reactivity can only go down. We
have seen that in our simulations.\cite{ned02}

We expect that our results will change only little when we change the
details of the reaction or when interaction between the particles are
included.\cite{sho97} Our results are a consequence of the non-passing
of the particles and of the fact that particles that stay longer in the
system have a higher probability of being converted. We therefore also
do not expect our results to change when we change our model of discrete
sites to some continuous model.

%
%

%
%

%
%
%

\begin{thebibliography}{16}
\expandafter\ifx\csname natexlab\endcsname\relax\def\natexlab#1{#1}\fi
\expandafter\ifx\csname bibnamefont\endcsname\relax
  \def\bibnamefont#1{#1}\fi
\expandafter\ifx\csname bibfnamefont\endcsname\relax
  \def\bibfnamefont#1{#1}\fi
\expandafter\ifx\csname citenamefont\endcsname\relax
  \def\citenamefont#1{#1}\fi
\expandafter\ifx\csname url\endcsname\relax
  \def\url#1{\texttt{#1}}\fi
\expandafter\ifx\csname urlprefix\endcsname\relax\def\urlprefix{URL }\fi
\providecommand{\bibinfo}[2]{#2}
\providecommand{\eprint}[2][]{\url{#2}}

\bibitem[{\citenamefont{Meier et~al.}(1996)\citenamefont{Meier, Olson, and
  Baerlocher}}]{mei96}
\bibinfo{author}{\bibfnamefont{W.~H.} \bibnamefont{Meier}},
  \bibinfo{author}{\bibfnamefont{D.~H.} \bibnamefont{Olson}}, \bibnamefont{and}
  \bibinfo{author}{\bibfnamefont{C.}~\bibnamefont{Baerlocher}},
  \emph{\bibinfo{title}{Atlas of Zeolite Structure Types}}
  (\bibinfo{publisher}{Elsevier}, \bibinfo{address}{London},
  \bibinfo{year}{1996}).

\bibitem[{\citenamefont{Privman}(1997)}]{pri97}
\bibinfo{author}{\bibfnamefont{V.}~\bibnamefont{Privman}},
  \emph{\bibinfo{title}{Nonequilibrium Statistical Mechanics in One Dimension}}
  (\bibinfo{publisher}{Cambridge University Press},
  \bibinfo{address}{Cambridge}, \bibinfo{year}{1997}).

\bibitem[{\citenamefont{Hahn and K\"arger}(1996)}]{hah96}
\bibinfo{author}{\bibfnamefont{K.}~\bibnamefont{Hahn}} \bibnamefont{and}
  \bibinfo{author}{\bibfnamefont{J.}~\bibnamefont{K\"arger}},
  \bibinfo{journal}{J.\ Phys.\ Chem.\/} \textbf{\bibinfo{volume}{100}},
  \bibinfo{pages}{316} (\bibinfo{year}{1996}).

\bibitem[{\citenamefont{Fedders}(1978)}]{fed78}
\bibinfo{author}{\bibfnamefont{P.~A.} \bibnamefont{Fedders}},
  \bibinfo{journal}{Phys.\ Rev.\ B\/} \textbf{\bibinfo{volume}{17}},
  \bibinfo{pages}{40} (\bibinfo{year}{1978}).

\bibitem[{\citenamefont{van Beijeren et~al.}(1983)\citenamefont{van Beijeren,
  Kehr, and Kutner}}]{bei83}
\bibinfo{author}{\bibfnamefont{H.}~\bibnamefont{van Beijeren}},
  \bibinfo{author}{\bibfnamefont{K.~W.} \bibnamefont{Kehr}}, \bibnamefont{and}
  \bibinfo{author}{\bibfnamefont{R.}~\bibnamefont{Kutner}},
  \bibinfo{journal}{Phys.\ Rev.\ B\/} \textbf{\bibinfo{volume}{28}},
  \bibinfo{pages}{5711} (\bibinfo{year}{1983}).

\bibitem[{\citenamefont{Keffer et~al.}(1995)\citenamefont{Keffer, McCormick,
  and Davis}}]{kef95}
\bibinfo{author}{\bibfnamefont{D.}~\bibnamefont{Keffer}},
  \bibinfo{author}{\bibfnamefont{A.~V.} \bibnamefont{McCormick}},
  \bibnamefont{and} \bibinfo{author}{\bibfnamefont{H.~T.} \bibnamefont{Davis}},
  in \emph{\bibinfo{booktitle}{Proceedings from the XI International Workshop
  on Condensated Matter Theories, Caracas}} (\bibinfo{year}{1995}).

\bibitem[{\citenamefont{Schuring et~al.}(2000)\citenamefont{Schuring, Jansen,
  and van Santen}}]{sch00d}
\bibinfo{author}{\bibfnamefont{D.}~\bibnamefont{Schuring}},
  \bibinfo{author}{\bibfnamefont{A.~P.~J.} \bibnamefont{Jansen}},
  \bibnamefont{and} \bibinfo{author}{\bibfnamefont{R.~A.} \bibnamefont{van
  Santen}}, \bibinfo{journal}{J.\ Phys.\ Chem.\ B\/}
  \textbf{\bibinfo{volume}{104}}, \bibinfo{pages}{941} (\bibinfo{year}{2000}).

\bibitem[{\citenamefont{R\"odenbeck et~al.}(1995)\citenamefont{R\"odenbeck,
  K\"arger, and Hahn}}]{rod95}
\bibinfo{author}{\bibfnamefont{C.}~\bibnamefont{R\"odenbeck}},
  \bibinfo{author}{\bibfnamefont{J.}~\bibnamefont{K\"arger}}, \bibnamefont{and}
  \bibinfo{author}{\bibfnamefont{K.}~\bibnamefont{Hahn}}, \bibinfo{journal}{J.\
  Catal.\/} \textbf{\bibinfo{volume}{157}}, \bibinfo{pages}{656}
  (\bibinfo{year}{1995}).

\bibitem[{\citenamefont{Nelson and Auerbach}(1999)}]{nel99}
\bibinfo{author}{\bibfnamefont{P.~H.} \bibnamefont{Nelson}} \bibnamefont{and}
  \bibinfo{author}{\bibfnamefont{S.~M.} \bibnamefont{Auerbach}},
  \bibinfo{journal}{J.\ Chem.\ Phys.\/} \textbf{\bibinfo{volume}{110}},
  \bibinfo{pages}{9235} (\bibinfo{year}{1999}).

\bibitem[{\citenamefont{Nedea et~al.}(2002)\citenamefont{Nedea, Jansen,
  Lukkien, and Hilbers}}]{ned02}
\bibinfo{author}{\bibfnamefont{S.~V.} \bibnamefont{Nedea}},
  \bibinfo{author}{\bibfnamefont{A.~P.~J.} \bibnamefont{Jansen}},
  \bibinfo{author}{\bibfnamefont{J.~J.} \bibnamefont{Lukkien}},
  \bibnamefont{and} \bibinfo{author}{\bibfnamefont{P.~A.~J.}
  \bibnamefont{Hilbers}}, \bibinfo{journal}{Phys.\ Rev. E\/}
  \textbf{\bibinfo{volume}{65}}, \bibinfo{pages}{066701}
  (\bibinfo{year}{2002}).

\bibitem[{\citenamefont{Krishna}(2000)}]{kri00}
\bibinfo{author}{\bibfnamefont{R.}~\bibnamefont{Krishna}},
  \bibinfo{journal}{Chem.\ Phys.\ Lett.\/} \textbf{\bibinfo{volume}{326}},
  \bibinfo{pages}{477} (\bibinfo{year}{2000}).

\bibitem[{\citenamefont{Tsikoyiannis and Wei}(1991)}]{tsi91}
\bibinfo{author}{\bibfnamefont{J.~G.} \bibnamefont{Tsikoyiannis}}
  \bibnamefont{and} \bibinfo{author}{\bibfnamefont{J.}~\bibnamefont{Wei}},
  \bibinfo{journal}{Chem.\ Eng.\ Sci.\/} \textbf{\bibinfo{volume}{46}},
  \bibinfo{pages}{233} (\bibinfo{year}{1991}).

\bibitem[{\citenamefont{K\"arger et~al.}(1992)\citenamefont{K\"arger, Petzold,
  Pfeiffer, Ernst, and Weitkamp}}]{kar92}
\bibinfo{author}{\bibfnamefont{J.}~\bibnamefont{K\"arger}},
  \bibinfo{author}{\bibfnamefont{M.}~\bibnamefont{Petzold}},
  \bibinfo{author}{\bibfnamefont{H.}~\bibnamefont{Pfeiffer}},
  \bibinfo{author}{\bibfnamefont{S.}~\bibnamefont{Ernst}}, \bibnamefont{and}
  \bibinfo{author}{\bibfnamefont{J.}~\bibnamefont{Weitkamp}},
  \bibinfo{journal}{J.\ Catal.\/} \textbf{\bibinfo{volume}{136}},
  \bibinfo{pages}{283} (\bibinfo{year}{1992}).

\bibitem[{\citenamefont{Gelten et~al.}(1999)\citenamefont{Gelten, van Santen,
  and Jansen}}]{gel99}
\bibinfo{author}{\bibfnamefont{R.~J.} \bibnamefont{Gelten}},
  \bibinfo{author}{\bibfnamefont{R.~A.} \bibnamefont{van Santen}},
  \bibnamefont{and} \bibinfo{author}{\bibfnamefont{A.~P.~J.}
  \bibnamefont{Jansen}}, in \emph{\bibinfo{booktitle}{Molecular Dynamics: From
  Classical to Quantum Methods}}, edited by
  \bibinfo{editor}{\bibfnamefont{P.~B.} \bibnamefont{Balbuena}}
  \bibnamefont{and} \bibinfo{editor}{\bibfnamefont{J.~M.}
  \bibnamefont{Seminario}} (\bibinfo{publisher}{Elsevier},
  \bibinfo{address}{Amsterdam}, \bibinfo{year}{1999}).

\bibitem[{\citenamefont{van Kampen}(1981)}]{kam81}
\bibinfo{author}{\bibfnamefont{N.~G.} \bibnamefont{van Kampen}},
  \emph{\bibinfo{title}{Stochastic Processes in Physics and Chemistry}}
  (\bibinfo{publisher}{North-Holland}, \bibinfo{address}{Amsterdam},
  \bibinfo{year}{1981}).

\bibitem[{\citenamefont{Sholl and Fichthorn}(1997)}]{sho97}
\bibinfo{author}{\bibfnamefont{D.~S.} \bibnamefont{Sholl}} \bibnamefont{and}
  \bibinfo{author}{\bibfnamefont{K.~A.} \bibnamefont{Fichthorn}},
  \bibinfo{journal}{Phys.\ Rev.\ Lett.\/} \textbf{\bibinfo{volume}{79}},
  \bibinfo{pages}{3569} (\bibinfo{year}{1997}).

\end{thebibliography}
\end{document}